\documentclass[structabstract]{aa}  
\usepackage{graphicx}
\usepackage{txfonts}
\usepackage{psfig}
\usepackage{natbib}
\usepackage[english]{babel}  
\usepackage{url}

\begin{document}
   \title{The shape of the cutoff in the synchrotron emission of SN 1006 observed with \emph{XMM-Newton}}

   \author{M. Miceli
          \inst{1}
          \and
          F. Bocchino\inst{1}
	  \and
	  A. Decourchelle\inst{2}
	  \and
	  J. Vink\inst{3}
	  \and
	  S. Broersen\inst{3}
          \and
	  S. Orlando\inst{1}
	  }

   \institute{INAF-Osservatorio Astronomico di Palermo, Piazza del Parlamento 1, 90134 Palermo, Italy
\\
              \email{miceli@astropa.unipa.it}
         \and
Service d'Astrophysique/IRFU/DSM, CEA Saclay, Gif-sur-Yvette, France
	 \and
Astronomical Institute ``Anton Pannekoek", University of Amsterdam, P.O. Box 94249, 1090 GE Amsterdam, The Netherlands	
             }

\date{}


  \abstract
   {Synchrotron X-ray emission from the rims of young supernova remnants allows us to study the high-energy tail of the electrons accelerated at the shock front.}
   {The analysis of X-ray spectra can provide information on the physical mechanisms that limit the energy achieved by the electrons in the acceleration process. We aim at verifying whether the maximum electron energy in SN 1006 is limited by synchrotron losses and at obtaining information on the shape of the cutoff in the X-ray synchrotron emission. }
   {We analyzed the deep observations of the \emph{XMM-Newton} SN~1006 Large Program.  We performed spatially resolved spectral analysis of a set of small regions in the nonthermal limbs and studied the X-ray spectra by adopting models that assume different electron spectra.}
   {We found out that a loss-limited model provides the best fit to all the spectra and this indicates that the shape of the cutoff in the electron momentum (p) distribution has the form $\rm{exp}[-(p/p_{cut})^2]$. We also detected residual thermal emission from shocked ambient medium and confirmed the reliability of previous estimates of the post-shock density.}
   {Our results indicate that radiative losses play a fundamental role in shaping the electron spectrum in SN 1006.}


   \maketitle
%

\section{Introduction}
\label{intro}

The shocks of supernova remnants (SNRs) are efficient sites of particle acceleration and are candidates for being the main source of the observed spectrum of cosmic rays up to at least $3\times10^{3}$ TeV \citep{be87,bv07}. 
The first evidence for shock acceleration of high energy electrons in SNRs has been obtained by detecting the non-thermal X-ray emission of SN~1006 \citep{kpg95}. Afterwords, X-ray synchrotron emission from high-energy electrons (up to TeV energies) has been observed in other young SNRs \citep{rey08,vin12}. Different physical mechanisms can be invoked to limit the maximum energy achieved by the electrons in the acceleration process, as for example, radiative losses, limited acceleration time available, change in the availability of MHD waves above some wavelength (i.~e. loss limited, time limited, and escape limited scenarios, see \citealt{rey08}).

The nonthermal emission in SN~1006 shows a characteristic bilateral morphology with two opposed (at northeast and southwest) radio, X-ray, and $\gamma-$ray bright limbs, though localized non-thermal X-ray emission has been observed also near the northwestern thermal limb, possibly associated with a fast ejecta knot \citep{bvm13}. The X-ray emission from the nonthermal limbs has been traditionally modelled as synchrotron radiation from a power law distribution of electron energies with an exponential cutoff (SRCUT model, \citealt{rk99}). This simple model provides a good description of the X-ray spectra extracted from regions in the northeastern and southwestern limbs (e. g., \citealt{rbd04}, \citealt{mbi09}, and \citealt{kpm10}, hereafter K10) and shows large-scale variation of the cutoff frequency, $\nu_{cut}$, which peaks at the nonthermal limbs and drops down (by more than a factor of ten) in thermal limbs and toward the center of the remnant (\citealt{rbd04}, \citealt{mbi09}).
More recent observations clearly unveiled the presence of small-scale variations of $\nu_{cut}$ that have been resolved down to the scales of the synchrotron filaments (K10, Decourchelle et al., in preparation).

K10 found a spatial correlation between the X-ray flux and the cutoff frequency in the northeastern limb of SN~1006 which may indicate that the cutoff frequency depends on the magnetic-field strength, $B$. In the loss limited scenario, the cutoff frequency does not depend on $B$ \citep{rey08}, therefore K10 argued that the maximum energy of accelerated electrons is not limited by synchrotron losses,  but by some other effect (i. e., it is time-limited or escape-limited, see \citealt{rey08}). Alternatively, one has to assume that particle acceleration depends on some other effects (e. g., the shock obliquity).

On the other hand, the effects of synchrotron losses on the shape of the electron (and photon) spectrum are expected to be relevant in SN~1006, as explained below. Different estimates of the downstream magnetic field, $B_s$ in the non-thermal limbs of SN~1006 show that $B_s\sim100~\mu$G. In particular, \citet{pmb06} derived $B_s\sim100~\mu$G and \citet{hvb12} derived $B_s\sim80~\mu$G from the thickness of the non-thermal X-ray filaments in the northeastern limb of SN 1006. \citet{mab10} compared the radial profile of the X-ray emission in this limb with their non-linear diffusive shock acceleration model, deriving $B_s\sim90~\mu$G. Moreover, \citet{bkv12} fitted the (global) SN 1006 multiwavelength nonthermal emission with their model, obtaining $B_s\sim150~\mu$G and a similar result was previously achieved by \citet{kbv05}. These values are therefore widely present in the literature and have been successfully adopted also to constrain the topology of the magnetic field around SN~1006 \citep{bom11}. The 
timescale of synchrotron cooling is 
\begin{equation}
t_{sync}=12.5E^{-1}_{100}B^{-2}_{100} ~{\rm ~ yr}
\end{equation}
where $E_{100}$ is the electron energy in units of 100 TeV and $B_{100}$ is the magnetic field in units of 100 $\mu$G \citep{lon94}. By equating $t_{sync}$ to the age of SN~1006 and putting $B_{100}=1$, we derive the electron energy $E^{*}_{100}=0.012$. The synchrotron emission of electrons with this energy peaks at
\begin{equation}
h \nu^{*}_{peak}=1.8\times10^4~E^{*2}_{100}B_{100}=2.7~ {\rm ~ eV}.
\end{equation}
At energies higher than $E^{*}_{100}$ the synchrotron losses make the spectrum steeper by one power of $E_{100}$ \citep{lon94}. Notice that, even by assuming a very low value of $B_s=40~\mu$G, we obtain $h\nu_{peak}=46$ eV. Therefore, significant synchrotron cooling for the electrons is expected over the lifetime of SN 1006.
However, to observe signatures of synchrotron cooling in the X-ray spectra, it is necessary that the acceleration time-scale, $t_{acc}$, is not much shorter than the synchrotron cooling time. We estimate the acceleration time-scale as $t_{acc}=3/(V_1-V_2)~(D_1/V_1+D_2/V_2)$ \citep{dru83}, where $V_{1,2}$ and $D_{1,2}$ are the upstream$/$downstream bulk velocities and diffusion coefficients. In the case of Bohm diffusion $D_{1,2}=p(1/3)c^2/(qB_{1,2})$, where $p$ and $q$ are the momentum and charge of electrons, respectively, and $c$ is the speed of light. We find that $t_{acc}\approx t_{sync}$ for electrons whose synchrotron emission peaks in the X-ray band. In particular, if we assume $B_2=100$ $\mu$G, and $B_1=B_2/\sqrt{11}$ (that is the increase of the isotropic random $B$ field for a shock compression ratio $r=4$) and $V_1=5000$ km$/$s (as measured by \citealt{kpl09}) and $V_2=1250$ km$/$s, we find that $t_{acc} = t_{sync}$ for electrons whose synchrotron emission peaks at 2 keV (and $t_{acc}\sim 0.5 t_
{sync}$ at 1 keV). We therefore conclude that signatures of a loss-dominated 
spectrum should be detected in the X-ray band.

The loss-limited scenario has been successfully adopted to describe the $Suzaku$ global (i. e. extracted from the whole SNR) spectrum of RX J1713.7-3946 (\citealt{za10}, \citealt{tua08}, \citealt{uat07}) and of Tycho \citep{mc12}. We here analyze X-ray spectra extracted from narrow and spectrally homogeneous regions at the nonthermal limbs of SN~1006, to discriminate between the loss-limited and the time-$/$escape-limited scenarios. In the loss-limited case, we expect a steepening of the electron spectrum (as explained before) and a different shape of the cutoff, which is not a simple exponential (as in the SRCUT model), but is $\propto \rm{exp}[-(E/E_{cut})^2]$, as derived by \citet{za07} and \citet{bla10}. Conversely, a constant spectral slope and an exponential cutoff would not be consistent with the loss-limited case and would support the findings of K10.

The paper is organized as follows: in Sect. \ref{data} we describe the data analysis procedure and the different spectral models, in Sect. \ref{Results}, we show the results of the spatially resolved spectral analysis; and, finally, we discuss our conclusions in Sect. \ref{Conclusions}.


\section{Data analysis and spectral models}
\label{data}

We analyze the data obtained within the \emph{XMM-Newton} Large Program of observations of SN~1006 (PI A. Decourchelle, 700 ks of total exposure time, hereafter LP observations) together with older \emph{XMM-Newton} observations (hereafter archive observations). The data reduction was performed with the Science Analysis System (SAS V12) by adopting the procedures described in detail in \citet{mbd12} (hereafter M12).

\begin{figure}[tb!]
 \centerline{\hbox{     
     \psfig{figure=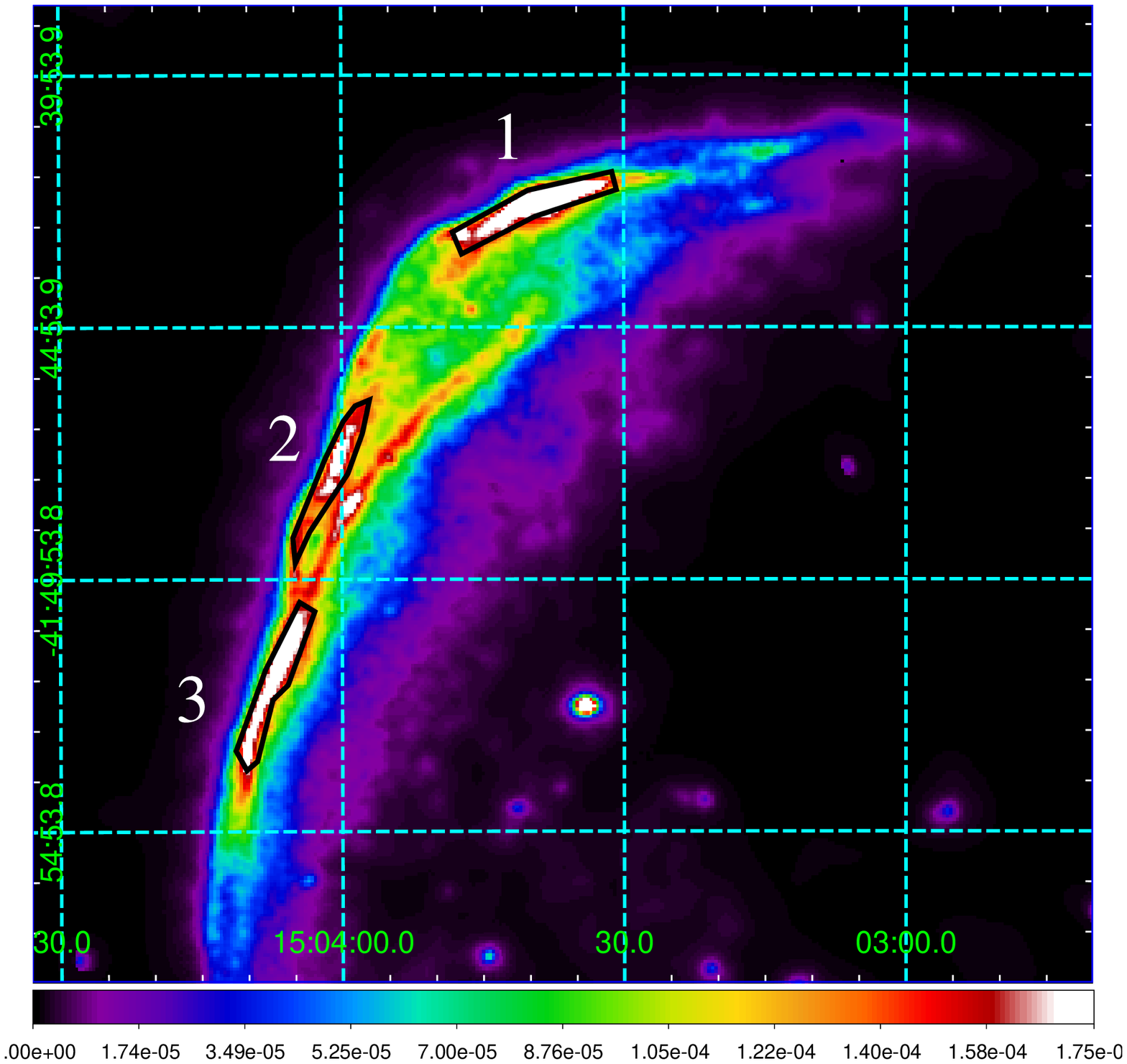,width=\columnwidth}}}
 \centerline{\hbox{     
     \psfig{figure=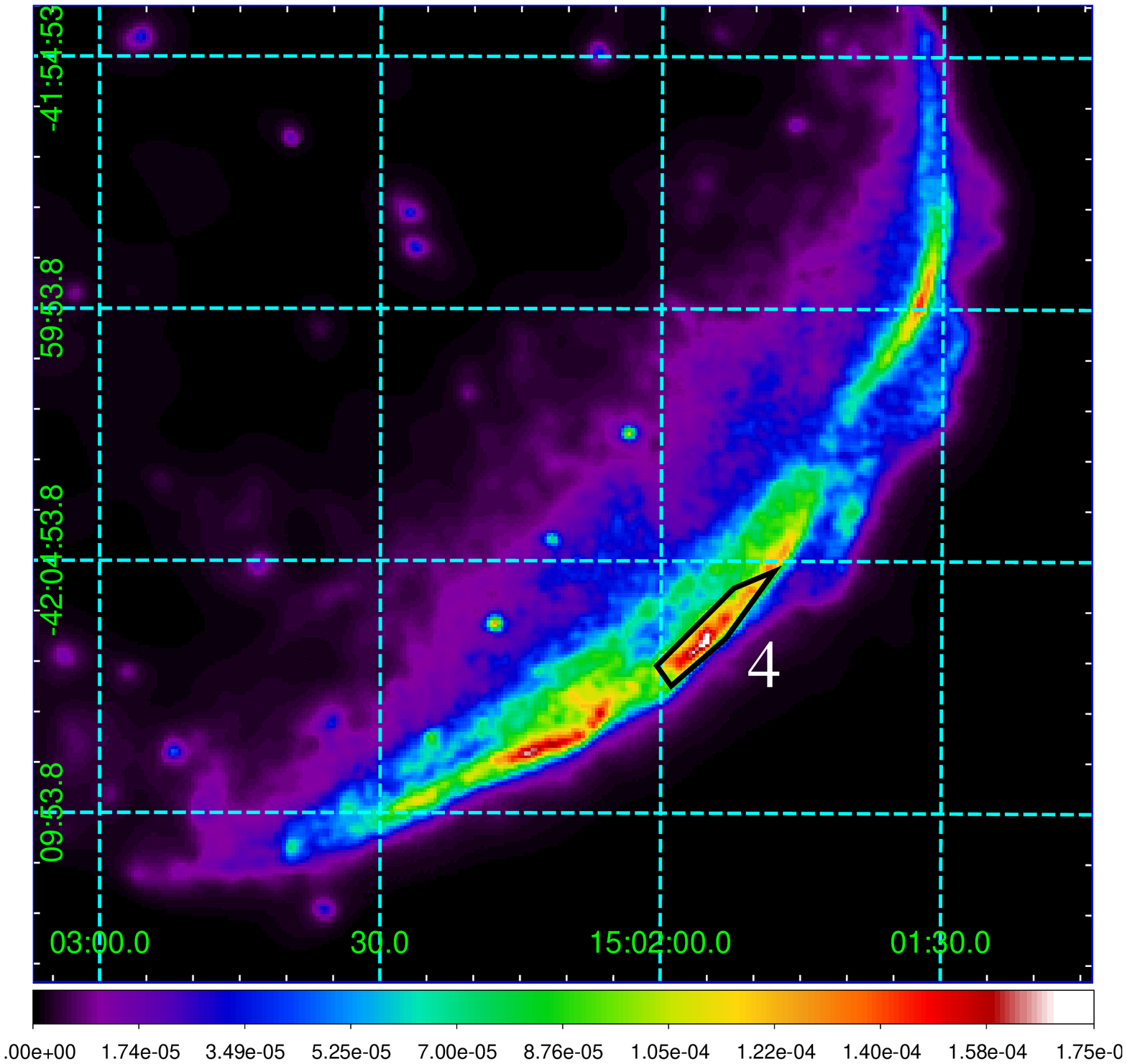,width=\columnwidth}}}
\caption{Mosaicked count-rate images (MOS-equivalent counts per second per bin) of the northeastern (\emph{upper panel}) and southwestern (\emph{lower panel}) limbs of SN~1006 in the $2-4.5$ keV band. The bin size is $4''$ and the image is adaptively smoothed to a signal-to-noise ratio ten. The regions selected for the spectral analysis of the rim are superimposed. North is up and East is to the left. }
\label{fig:regions}
\end{figure}

We focus on regions of the shell where the X-ray emission is dominated by synchrotron radiation and the contribution of the thermal component is negligible (see Sect. \ref{thermal}). For our study, it is important to analyze spectrally homogeneous regions, so we selected relatively small regions to minimize variations in the cutoff frequency. The regions selected for our spatially resolved spectral analysis are shown in Fig. \ref{fig:regions} and extend approximately 3.5' azimuthally and 0.5' radially (the cutoff frequency varies exponentially in the radial direction, as shown by \citealt{rbd04}, so it is important to minimize the radial width). We selected three regions (regions $1-3$ in Fig. \ref{fig:regions}) in the bright northeastern limb, while in the fainter southeastern limb there is only one region (region 4) where it is possible to achieve statistically significant results (our criterium is to collect at least $13000$ photon counts in the $0.5-7.5$ keV band).

The source regions $1-3$ are covered by the observations ID 0555630201, 0555630301 (LP observations), 0143980201, and 0077340201 (archive observations), while region $4$ is covered by observation ID 0653860101 (LP observations) and 0202590101 (archive observations). The screened MOS1$/$MOS2$/$pn exposure times are $91/96/64$ ks, $91/92/77$ ks, $16/17/12$ ks, $25/26/21$ ks,  $103/105/88$ ks, and $27/29/19$ ks for observations 0555630201, 0555630301, 0143980201, 0077340201, 0653860101, and 0202590101 respectively. MOS1,2 and pn spectra of the different observations were fitted simultaneously.

Regions $1-4$ were chosen to maximize the photon counts, because the two spectral models we want to test are somehow similar in the X-ray band. In particular, the SRCUT model assumes that the energy spectrum of electrons follows the relation 
\begin{equation}
N(E)\propto E^{-s}\rm{exp}~(-E/E_{cut})
\end{equation}
and \citet{pbm09} showed that the corresponding synchrotron spectrum can be well approximated\footnote{In this case, the $\delta-$function approximation of the single electron emissivity does not provide an accurate description of the spectrum, as pointed-out by \citet{rk99}.} by the formula:
\begin{equation}
S_X^{sr}\propto\nu^{-(s+1)/2}\rm{exp}~[-\beta ( h\nu/ h\nu_{cut})^{0.364}]
\label{srcut}
\end{equation}
where $\beta=1.46+0.15(2-s)$. On the other hand, \citet{za07} have shown that the electron spectrum at the shock in the loss-dominated case is\footnote{In the loss-limited scenario we will adopt the notation $E_0$ and $h\nu_0$ for the cutoff energy of electrons and photons, respectively.} 
\begin{equation}
N(E)\propto E^{-2}[1+a(E/E_0)^b]^c \rm{exp}~[-(E/E_0)^2]
\end{equation}
where $a=0.66~(0.523)$, $b=5/2~(9/4)$, $c=9/5~(2)$, depending whether the magnetic field downstream is compressed by a factor $\kappa=1~(\sqrt{11})$ with respect to upstream. The resulting X-ray synchrotron spectrum integrated over the downstream region (the upstream contribution being negligible, see Fig. 5 in \citealt{za07}) has the form
\begin{equation}
S_X^{ll}\propto h\nu^{-2}[1+l(h\nu/h\nu_0)^m]^n\rm{exp}~(-\sqrt{h\nu/h\nu_{0}})
\label{zira}
\end{equation}
where $l=0.46~(0.38)$, $m=0.6~(0.5)$, $n=11/4.8~(11/4)$ for $\kappa=1~(\sqrt{11})$. 

The shape of the cutoff in the spectrum of Eq. (\ref{srcut})  is slightly steeper than that expected in the loss-limited case described by Eq. (\ref{zira}). Another notable difference between the two spectra is the presence of a power law term in Eq. (\ref{zira}). This term can play an important role at energies $\la h\nu_0$. 

We performed spectral analysis in the $0.5-7.5$ keV energy band, by using XSPEC V12, where we introduced the loss-limited model by following Eq. (\ref{zira}). We modelled the interstellar absorption with the TBABS model, and we set the absorbing column to $N_H=7\times10^{20}$ cm$^{-2}$, in agreement with \citet{dgg02} (but see Sect. \ref{nont}).

In the SRCUT model, we fixed the normalization to the value derived from the radio image \citep{mbi09} in the corresponding region (but see also Sect. \ref{nont}). The photon index $\alpha=(s-1)/2$ has been measured from the radio spectrum for the entire remnant by \citet{ahs08}, who found $\alpha=0.60^{+0.08}_{-0.09}$, in agreement with the value $\alpha=0.6$ reported by \citet{gre09}. The global radio emission is dominated by the nonthermal limbs (where our regions are located). Moreover, there is little evidence of spatial variability in the radio spectrum of SN 1006 and the spectral slopes of two nonthermal limbs are consistent with being the same (see \citealt{ahs08} and references therein). Therefore, we can assume that $\alpha\sim0.6$ is appropriate for our regions. However, X-ray spectral analysis performed in selected regions of the nonthermal limbs has shown that a value $\alpha=0.5$ (\citealt{mbi09}, K10) should be preferred. We therefore let $\alpha$ free to vary in 
the range $0.5-0.7$. The cutoff frequency is a free parameter in the fitting procedure.

In the loss limited model, the only free parameters are the normalization and the cutoff energy $h\nu_0$. We found out that it is not possible to discriminate between the cases $\kappa=1$ and $\kappa=\sqrt{11}$, though the latter provides slightly better fits to the spectra. Nevertheless, the improvements in the $\chi^2$ values are minimal (on average, $\Delta\chi^2\sim30$ with $\sim 3000$ d. o. f.) and it is not possible to rule out the other scenario. In the following, we will show only the results obtained with $\kappa=\sqrt{11}$.

The spectra of regions $1-4$ are dominated by synchrotron emission, but some contribution from thermal emission is visible as residuals at $\sim0.57$ keV and $\sim0.66$ keV (i.~e. at the energies of the K-shell line complexes of O VII and O VIII). We adopted three different approaches to model this contribution, namely we included in the model: i) two narrow Gaussians; ii) a thermal component with the same parameters as the ejecta component of M12; iii) a thermal component with the same parameters as the shocked interstellar medium (ISM) component of M12 (see Sect. \ref{thermal} for further details). We verified that, in all the cases, there are no pronounced variations in the best-fit parameters of the non-thermal component and in the $\chi^2$ values. Therefore, in the next section, we report only the results obtained by adopting approach i).

\section{Results}
\label{Results}

\subsection{Nonthermal emission}
\label{nont}

\begin{figure}[tb!]
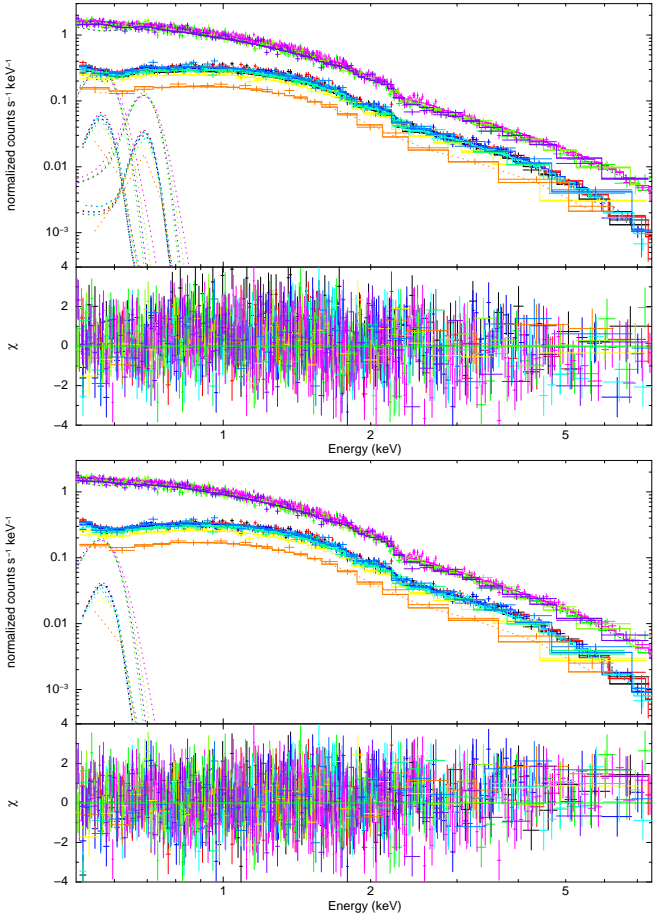

 \centerline{\hbox{     
     \psfig{figure=reg1_srcut_reb10.ps,angle=-90,width=\columnwidth}}}
 \centerline{\hbox{     
     \psfig{figure=reg1_zira_reb10.ps,angle=-90,width=\columnwidth}}}  
\caption{\emph{Upper panel}: PN (upper) and MOS (lower) spectra of region $1$ (shown in Fig. \ref{fig:regions}) with the corresponding SRCUT$+$Gaussians best-fit model and residuals (see Table \ref{tab:res}). The contribution of each components is shown. \emph{Lower panel}: same as upper panel for the loss limited$+$Gaussians model.}
\label{fig:spec}
\end{figure}

Table \ref{tab:res} shows the best-fit results obtained in the spectral regions of Fig. \ref{fig:regions} by adopting the SRCUT model and the loss limited model. Though both models provide good fits to the spectra, we found that \emph{in all the regions} the loss limited model describes the observed emission much better than the SRCUT model. 
As a representative case, Fig. \ref{fig:spec} shows the EPIC spectra extracted from region 1 with the corresponding SRCUT (plus two Gaussians component, upper panel) and loss limited (with only one Gaussian component, lower panel) models and residuals. 

The SRCUT model slightly underpredicts the low-energy part of the spectra (below $\sim1$ keV) and, as shown in Table \ref{tab:res}, requires the introduction of additional Gaussian components (and$/$or higher normalizations for these components) with respect to the loss limited model that, even with more degrees of freedom, provides lower values of the $\chi^2$. Let us consider, for example, region 1, that is the one with the highest surface brightness and the best statistics. In this region, if we remove the O VIII line complex (i. e. the gaussian at $0.69\pm0.01$ keV in Table \ref{tab:res}) from the SRCUT+Gaussians model, we obtain $\chi^2=4255.0$ with 3632 d. o. f., to be compared with the much smaller $\chi^2=3791.1$ obtained with the same number of d. o. f. in the loss limited scenario.
In principle, these problems at low energies may be caused by the fact that we fixed $N_H=7\times10^{20}$ cm$^{-2}$ in all the regions. We relaxed this assumption by letting the $N_H$ free to vary in the fitting. We found that, by adopting the loss limited model, we obtain values of $N_H$ consistent with $7\times10^{20}$ cm$^{-2}$ (at the $90\%$ confidence level) and no significant improvements in the quality of the fits. By adopting the SRCUT model, instead, we obtain new $\chi^2$ minima, namely $\chi^2=3941$ with 3631 d.~o.~f., $\chi^2=3139$ with 3017 d.~o.~f., $\chi^2=3749$ with 3284 d.~o.~f., and $\chi^2=1655$ with 1511 d.~o.~f. in regions $1-4$, respectively. These values are still much higher than those obtained in the loss limited scenario (see Table \ref{tab:res}). Moreover, the best fit values of the absorbing column in the SRCUT framework are rather low: $N_H=4.8\pm0.2\times10^{20}$ cm$^{-2}$, $N_H=5.8\pm0.1\times10^{20}$, $N_H=2.7\pm0.2\times10^{20}$ cm$^{-2}$, and $N_H=2.5^{+0.2}_{-0.3}\times10^{
20}$ cm$^{-2}$, in regions $1-4$, respectively. In particular, the values obtained in region 3 and 4 are unrealistically low and we can consider these variations in the $N_H$ as artifacts induced by the bad description of the low-energy part of the spectra provided by the SRCUT model. We therefore conclude that the loss limited model provides a better description of all the spectra.

\begin{center}
\begin{table*}
\begin{center}
\caption{Best-fit parameters (all errors at $90\%$ confidence level).}
\begin{tabular}{@{}lccccc} 
\hline\hline
     Parameters         &       Region 1         &     Region 2           &      Region 3       &  Region 4   \\ \hline
{\bf SRCUT+Gaussians}   &                        &                        &                     &             \\ 
    $\alpha$            &    $0.5000^{+0.001}$   &   $0.5000^{+0.001}$    &  $0.5000^{+0.001}$  & $0.5000^{+0.001}$   \\
$h\nu_{cut}$ (keV)     &$0.525^{0.002}_{-0.003}$ &$0.490^{+0.002}_{-0.003}$ &  $0.386\pm0.002$  & $0.423\pm0.003$ \\
Line E1 (keV)          &     $0.565\pm0.004$     &     $0.56\pm0.01$     &$0.569^{+0.001}_{-0.002}$& $0.572^{+0.003}_{-0.004}$ \\
Line Norm1 (cm$^{-2}$/s)&$8.7\pm0.7\times10^{-5}$&$2.1\pm0.5\times10^{-5}$&$2.25\pm0.08\times10^{-5}$&$1.24\pm0.07\times10^{-5}$\\
Line E2 (keV)           &      $0.69\pm0.01$     &       $-^{*}$          &  $0.685\pm0.006$    &  $0.69\pm0.01$ \\
Line Norm2 (cm$^{-2}$/s)&$2.0\pm0.3\times10^{-5}$&      $-^{*}$      &$6.1\pm0.4\times10^{-5}$&$3.2^{+0.4}_{-0.3}\times10^{-5}$\\ 
$\chi^2$ (d. o. f.)     &    $4132.1~(3630)$     &  $3187.5~(3016)$        & $4505.6~(3285)$      & $2103.9~(1512)$ \\ \hline
{\bf Loss limited+Gaussians}&                     &                       &                     &                   \\
$h\nu_0$ (keV)          &  $0.448\pm0.009$       &    $0.44\pm0.01$       &   $0.289\pm0.005$   & $0.303\pm0.008$ \\
Line E1 (keV)           &   $0.567\pm0.005$      &       $-^{*}$          &   $0.569\pm0.002$   &$0.569^{+0.004}_{-0.005}$ \\
Line Norm1 (cm$^{-2}$/s)&$6.0\pm0.8\times10^{-5}$ &        $-^{*}$     &$1.88\pm0.09\times10^{-5}$&$9.0\pm0.8\times10^{-5}$\\
Line E2 (keV)           &     $-^{*}$             &     $-^{*}$           & $0.67\pm0.01$ & $0.65^{+0.02}_{-0.01}$ \\
Line Norm2 (cm$^{-2}$/s)&     $-^{*}$           &     $-^{*}$        &$3.6^{+0.5}_{-0.4}\times10^{-5}$&$2.2\pm0.5\times10^{-5}$\\ 
$\chi^2$ (d. o. f.)     &   $3791.1~(3632)$       &   $3058.1~(3018)$     & $3496.1~(3285)$ &  $1554.8~(1512)$ \\ \hline 
\multicolumn{5}{l}{\footnotesize{$^*$ Model components with normalization consistent with zero at three sigmas were not included.}}\\
\label{tab:res}
\end{tabular}
\end{center}
\end{table*}
\end{center}

\begin{figure}[tb!]
 \centerline{\hbox{     
     \psfig{figure=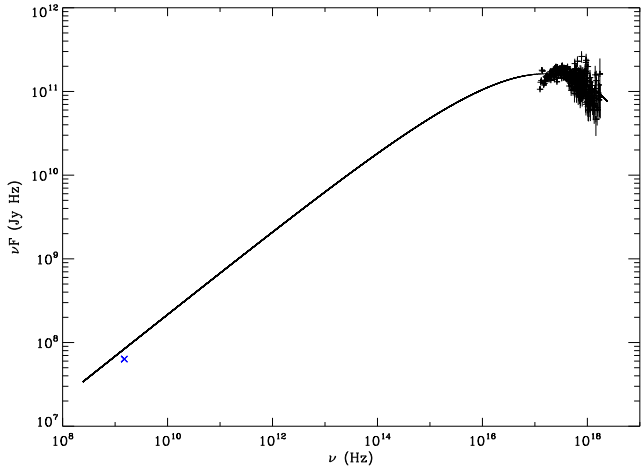,angle=90,width=\columnwidth}}}
\caption{MOS1 X-ray spectrum of region 1 (black crosses) with its best-fit model obtained by letting the SRCUT normalization free to vary (see Sect. \ref{nont}). The corresponding radio point is shown in blue.}
\label{fig:radiox}
\end{figure}

In the SRCUT scenario, $\alpha$ holds its lower limit ($0.5$) in all the spectral regions. This value is systematically lower than the best-fit radio spectral index ($0.6$) and this may indicate that the synchrotron spectrum flattens with increasing energy. A similar result has been obtained by \citet{ahs08} and has been interpreted as an evidence of a curved electron spectrum. If this is the case, the relatively poor fits that we obtained with the SRCUT model can be the result of a change in the spectral slope not accounted for in our model. We investigate this possibility by letting the normalization free in the SRCUT model (i. e. by not anchoring the spectrum to the corresponding value derived from the radio image). In this case, since the SRCUT model only describes the X-ray part of the spectrum (with spectral slope $\alpha=0.5$) and we do not include any change in the slope, we expect that by extrapolating the best-fit model to the radio band (where $\alpha=0.6$), we would obtain a radio flux that is 
much smaller than that observed. Indeed this is not the case, though the model provides very good fits, as shown below. By letting the normalization of the SRCUT component free, the $\chi^2$ values become in fact comparable to that obtained in the loss limited scenario, namely $\chi^2=3785.8$ (with 3636 d. o. f.), $\chi^2=3060.6$ (with 3020 d. o. f.), $\chi^2=3478.5$ (with 3291 d. o. f.), and $\chi^2=1599.2$ (with 1511 d. o. f.) in regions 1, 2, 3, and 4 respectively\footnote{The normalizations of the Gaussian components are also consistent with that obtained in the loss limited scenario and shown in Table 1}. However, the best fit model does not underpredict the radio flux. Instead, it systematically overpredicts the radio flux $F_{radio}$ at 1.5 GHz \emph{in all the spectral regions}. As an example, Fig. \ref{fig:radiox} shows the MOS1 X-ray spectrum of region 1 (black crosses) with its best-fit model obtained by letting the SRCUT normalization free to vary. The figure shows that the radio flux predicted 
by this model is higher than that observed (indicated by the blue cross in the figure), while, if the spectrum were steeper at lower energy, we would have expected the model to underpredict the radio flux. 
In particular, we obtain $F_{radio}=6.9^{+0.3}_{-0.1}\times10^{-2}$ Jy (to be compared with the observed value $F_{radio}=5.4\times10^{-2}$ Jy) in region 1 (see Fig. \ref{fig:radiox}), $F_{radio}=4.1^{+0.4}_{-0.1}\times10^{-2}$ Jy (to be compared with the observed value $F_{radio}=3.5\times10^{-2}$ Jy) in region 2, $F_{radio}=1.01\pm0.02\times10^{-1}$ Jy (to be compared with the observed value $F_{radio}=6.6\times10^{-2}$ Jy) in region 3, and $F_{radio}=6.0^{+0.1}_{-0.2}\times10^{-2}$ Jy (to be compared with the observed value $F_{radio}=4.0\times10^{-2}$ Jy) in region 4. This is at odds with expectations and shows that, even in the curved spectrum scenario, the SRCUT model does not provide a correct description of the spectra.

We conclude that it is not possible to adopt the SRCUT model, even if we invoke a curvature in the spectra. Instead, all the spectra can be naturally described within the loss-limited framework. As a cross-check, we calculated the energy $h\nu_{roll}$ at which the extrapolated radio spectrum intercept the loss limited model. This is the energy at which the electron spectrum becomes steeper by one power of $E$. Therefore, $h\nu_{roll}$ can be considered as a (rough) estimate of the peak energy $h\nu_{peak}$ for the electrons whose radiative loss time is comparable to the SN 1006 age (e. g. \citealt{mc12}). In our regions we found $h\nu_{peak}\sim50$ eV that, by using Eq. (1) and Eq. (2), indicates $B\sim40~\mu$G. Though this estimate of the downstream magnetic field is quite rough, it has the same order of magnitude than the more accurate values $B\sim100~\mu$G discussed in Sect. \ref{intro} and this indicates that our results are sound. We also point out that most of the $B$ estimates presented in Sect. \ref{intro} were obtained by assuming that the spectrum is loss-limited.

\subsection{Thermal emission}
\label{thermal}

As explained in Sect. \ref{nont}, a pure nonthermal model leaves pronounced residuals at the energies of the O VII and O VIII line complexes. This line emission originates in the ejecta and$/$or in the shocked ISM but, since the spectrum is synchrotron dominated, it is not possible to ascertain detailed information on the nature of the thermal emission. However, region $1$ is located where the distance between the shock front and the contact discontinuity is close to its maximum (see Fig. 4 and Fig. 6 in \citealt{mbi09}) and a shock breakout is present. In region $1$, therefore, it is quite unlikely to observe X-ray emission from shocked ejecta. We can then reasonably assume that the OVII line emission visible in the spectrum of region 1 (see Fig. \ref{fig:spec}) is associated with the shocked ISM only. This makes region 1 particularly interesting, since in other parts of the shell both ejecta and ISM contribute to the oxygen line emission.
In the loss limited model of Table 1, we replaced the Gaussian component at $0.567$ keV with a thermal PSHOCK model (\citealt{blr01}) having the same parameters as the ISM component of M12 (we let only the emission measure, $EM$, free to vary in the fitting process). This operation does not alter the best-fit value of the loss limited model that are consistent within less than one sigma with those reported in Table 1 (this new model gives $\chi^2=3786.6$, with 3633 d. o. f.). From the best fit value of $EM$, we derived the density as in M12 and we obtained $n_{ISM}=0.29\pm0.02$ cm$^{-3}$. Though it was not possible to accurately constrain the thermal spectrum, this value is in remarkable agreement with the characteristic post-shock density of the ISM derived by M12 in the southeastern thermal limb.
To confirm our detection of thermal X-ray emission from shocked ISM in the northeastern nonthermal limb, we also excluded a possible ejecta origin for the line emission.
In fact, if we assume that the line emission originates in the ejecta (i. e. we introduce a thermal VPSHOCK model with the same parameters as the ejecta component of M12) we derive an ejecta density $n_{ej}=0.053\pm0.003$ cm$^{-3}$. This value is one order of magnitude smaller than the characteristic ejecta density in SN 1006 (e. g. M12). This confirms that the line emission detected in region 1 originates in the shocked ISM.

In Sect. \ref{nont}, we have shown that the adoption of the SRCUT model to describe the non-thermal emission can introduce some bias in the determination of the thermal emission. This is because the SRCUT model slightly underestimates the low energy non-thermal flux, thus determining a possible overestimation of the (soft) thermal contribution.
M12 found that the post-shock density significantly increases near the nonthermal limbs, thus indicating the presence of shock modification induced by hadron acceleration at the shock front. The significance of the detection of the ISM component drops down as the synchrotron contribution increases (M12), thus indicating that the additional ISM component has indeed a thermal origin and is not an artifact due to a misdescription of the synchrotron emission (in this case we should have detected it with higher statistical significance in synchrotron-dominated regions).
Nevertheless, since M12 adopted the SRCUT scenario, it is important to verify that, by modeling the non-thermal emission with the loss limited model, the best fit parameters of the thermal components do not change significantly. We therefore fitted the spectra of regions $a-h$ of M12 by substituting the SRCUT component with the loss limited model. We found that in these regions the results do not change and, in particular, the post-shock density of the ISM component is not affected by this issue. For example, the updated values of the ISM density are consistent within less than one sigma in region $a$ and within less than 0.5 sigma in region $b$ with those reported in M12 (the discrepancies are even smaller in the other regions). We therefore confirm the results obtained in M12.

\section{Discussion and conclusions}
\label{Conclusions}

We have analyzed a set of deep \emph{XMM-Newton} observations of SN~1006 that have allowed us to study the shape of the cutoff of the synchrotron emission in the nonthermal limbs.
We found that the SRCUT model does not describe correctly the observed spectra and that the loss limited model developed by \citet{za07} provides a much better description of the cutoff region. 
The estimates of the downstream magnetic field also concur in indicating that the loss limited mechanism is at work in SN 1006.

K10 found that there is a spatial correlation between the X-ray flux and the cutoff frequency in the northeastern limb of SN~1006. This result may indicate that the cutoff frequency depends on the magnetic-field strength, at odds with what happens if the maximum energy of the accelerated electron is loss limited. Nevertheless, they suggest that this result can be consistent with the loss limited scenario if the rate of particle injection and$/$or acceleration depends on some effect not yet accounted for, as, for example, the shock obliquity. Therefore, the findings of K10 do not rule out the loss limited model and are not in contradiction with our results.

We also obtained evidence for the presence of thermal X-ray emission from shocked ISM in the northeastern limb of SN 1006 and verified that the ISM density values presented in M12 are reliable. 

M12 found indications for the presence of shock modification induced by hadron acceleration in the nonthermal limbs of SN 1006. We point out that the loss limited model adopted here does not take into account non-linear effects caused by hadrons on the shock structure. A preliminary model developed by \citet{bla10} shows that, in modified shocks, the spectra in the presence of synchrotron losses somehow differ from the test particle case. Also, the model adopted here assumes Bohm diffusion, but, in general, the shape of the cutoff depends on the nature of the diffusion process (e. g. Kolmogorov diffusion, diffusion constant in momentum), as shown by \citet{bla10}.

The available data do not allow us to perform more accurate diagnostics. However, it will be possible to discriminate between the aforementioned scenarios with the next generation of X-ray telescopes. We performed spectral simulations to synthesize focal plane spectra observed with the WFI camera of the proposed Athena+ mission\footnote{The Athena+ response files are available at http://www.the-athena-x-ray-observatory.eu} and verified that, for example, it will be possible to discriminate between the cases $\kappa=1,~\sqrt{11}$ through global fittings in the $0.3-10$ keV band with a single 80 ks observation of region 1. 


\begin{acknowledgements}
This paper was partially funded by the ASI-INAF contract I$/$009$/$10$/$0.
\end{acknowledgements}

\bibliographystyle{aa}





\end{document}